\documentclass[12pt]{article}
\usepackage{amsmath,amsfonts,amssymb,amsthm,amstext,amscd,eucal,xcolor}
\usepackage[all]{xy}
\usepackage{hyperref}
\usepackage{epsfig}
\usepackage{color}
\usepackage{graphicx}
\usepackage[active]{srcltx}
\makeatletter \@addtoreset{equation}{section}

\makeatletter\renewcommand\section{\@startsection {section}{1}{\z@}%
                                   {-3.5ex \@plus -1ex \@minus -.2ex}
                                   {2.3ex \@plus.2ex}%
                                   {\normalfont\large\bfseries}}
\renewcommand\subsection{\@startsection{subsection}{2}{\z@}%
                                     {-3.25ex\@plus -1ex \@minus -.2ex}%
                                     {1.5ex \@plus .2ex}%
                                     {\normalfont\bfseries}}

\begin{document}
\baselineskip 18pt%
\newcommand{\beq}{\begin{equation}}
\newcommand{\eeq}{\end{equation}}
\newcommand{\beqa}{\begin{eqnarray}}
\newcommand{\eeqa}{\end{eqnarray}}
\newcommand{\beqar}{\begin{eqnarray*}}
\newcommand{\eeqar}{\end{eqnarray*}}
\newcommand{\al}{\alpha}
\newcommand{\be}{\beta}
\newcommand{\del}{\delta}
\newcommand{\D}{\Delta}
\newcommand{\eps}{\epsilon}
\newcommand{\ga}{\gamma}
\newcommand{\Ga}{\Gamma}
\newcommand{\ka}{\kappa}
\newcommand{\inn}{\!\cdot\!}
\newcommand{\h}{\eta}
\newcommand{\kk}{\varphi}
\newcommand\F{{}_3F_2}
\newcommand{\la}{\lambda}
\newcommand{\La}{\Lambda}
\newcommand{\na}{\nabla}
\newcommand{\Om}{\Omega}
\newcommand{\p}{\phi}
\newcommand{\sig}{\sigma}
\renewcommand{\t}{\theta}
\newcommand{\z}{\zeta}
\newcommand{\ssc}{\scriptscriptstyle}
\newcommand{\eg}{{\it e.g.,}\ }
\newcommand{\ie}{{\it i.e.,}\ }
\newcommand{\labell}[1]{\label{#1}} 
\newcommand{\reef}[1]{(\ref{#1})}
\newcommand\prt{\partial}
\newcommand\veps{\varepsilon}
\newcommand\ls{\ell_s}
\newcommand\cF{{\cal F}}
\newcommand\cA{{\cal A}}
\newcommand\cS{{\cal S}}
\newcommand\cT{{\cal T}}
\newcommand\cC{{\cal C}}
\newcommand\cL{{\cal L}}
\newcommand\cG{{\cal G}}
\newcommand\cI{{\cal I}}
\newcommand\cl{{\iota}}
\newcommand\cP{{\cal P}}
\newcommand\cV{{\cal V}}
\newcommand\cg{{\it g}}
\newcommand\cR{{\cal R}}
\newcommand\cB{{\cal B}}
\newcommand\cO{{\cal O}}
\newcommand\tcO{{\tilde {{\cal O}}}}
\newcommand\bz{\bar{z}}
\newcommand\bw{\bar{w}}
\newcommand\hF{\hat{F}}
\newcommand\hA{\hat{A}}
\newcommand\hT{\hat{T}}
\newcommand\htau{\hat{\tau}}
\newcommand\hD{\hat{D}}
\newcommand\hf{\hat{f}}
\newcommand\hg{\hat{g}}
\newcommand\hp{\hat{\phi}}
\newcommand\hi{\hat{i}}
\newcommand\ha{\hat{a}}
\newcommand\hQ{\hat{Q}}
\newcommand\hP{\hat{\Phi}}
\newcommand\hS{\hat{S}}
\newcommand\hX{\hat{X}}
\newcommand\tL{\tilde{\cal L}}
\newcommand\hL{\hat{\cal L}}
\newcommand\tG{{\tilde G}}
\newcommand\tg{{\widetilde g}}
\newcommand\tphi{{\widetilde \phi}}
\newcommand\tPhi{{\widetilde \Phi}}
\newcommand\ta{{\tilde a}}
\newcommand\tb{{\tilde b}}
\newcommand\tf{{\tilde f}}
\newcommand\tF{{\widetilde F}}
\newcommand\tK{{\widetilde K}}
\newcommand\tR{{\tilde R}}
\newcommand\tpsi{{\tilde \psi}}
\newcommand\tX{{\widetilde X}}
\newcommand\tD{{\widetilde D}}
\newcommand\tO{{\widetilde O}}
\newcommand\tS{{\tilde S}}
\newcommand\tB{{\widetilde B}}
\newcommand\tA{{\widetilde A}}
\newcommand\tT{{\widetilde T}}
\newcommand\tC{{\widetilde C}}
\newcommand\tV{{\widetilde V}}
\newcommand\thF{{\widetilde {\hat {F}}}}
\newcommand\Tr{{\rm Tr}}
\newcommand\tr{{\rm tr}}
\newcommand\STr{{\rm STr}}
\newcommand\M[2]{M^{#1}{}_{#2}}
\parskip 0.3cm

\vspace*{1cm}
\begin{center}
{\Large {\bf  T-duality of D-brane  action  at order  $\alpha'$ \\ in bosonic string theory } }

\vspace*{15mm} 
\vspace*{1mm} 
{\large   Mohammad R. Garousi\footnote{garousi@um.ac.ir}, Ahmad Ghodsi\footnote{a-ghodsi@um.ac.ir},\\ Tooraj Houri\footnote{to$\_$ho296@stu-mail.um.ac.ir} and Ghadir Jafari\footnote{ghadir.jafari@stu-mail.um.ac.ir}}
\vspace*{1cm}

{ {Department of Physics, Ferdowsi University of Mashhad,\\
 P.O. Box 1436, Mashhad, Iran}} \\

\vskip 0.6 cm

\vspace*{1cm}
\end{center}

\begin{abstract}

 In bosonic string theory, it is known that the  Buscher rules for the T-duality transformations receive quantum corrections at order $\alpha'$. In this paper, we use the consistency of the gravity couplings on the D-brane effective action at order $\alpha'$, with the above T-duality transformations to find the B-field and the dilaton couplings. We show that these couplings are fully consistent with the corresponding disk-level S-matrix elements in string theory.

\end{abstract}
\newpage 
\section{Introduction  and results}

One of the most fantastic dualities of string theory is T-duality \cite{Kikkawa:1984cp,TB,Giveon:1994fu,Alvarez:1994dn}. It relates the bosonic string theory compactified on a circle with radius $\rho$ to the same theory compactified on another circle with radius $\alpha'/\rho$. It also relates the D$_p$-brane of the theory to the D$_{p-1}$-brane and  D$_{p+1}$-brane, depending on whether the original D$_p$-brane is along or orthogonal to the circle, respectively. The bulk  action $S$ which is given by the closed string field theory,  must be invariant under the T-duality  \cite{Kugo:1992md,Zwiebach:1992ie}, and the D$_p$-brane  action $S_{D_p}$ which is given by the  cubic string field theory \cite{Witten:1985cc},  must be covariant under the T-duality, \ie
\beqa
S\stackrel{T}{\longrightarrow} S\,, \qquad S_{D_p}\stackrel{T}{\longrightarrow}S_{D_{p\pm 1}}\,.
\eeqa
An appropriate framework  for incorporating the T-duality into these  actions   is   the double field theory formalism in which the fields depend both on the usual spacetime coordinates and on the winding coordinates \cite{Hull:2009mi,Hohm:2010jy,Hohm:2010pp}.  These actions can be expanded at low energy, \ie
\beqa
S\,=\,\sum_{n=0}^{\infty} (\alpha')^n S^{(n)}\,,\qquad S_{D_p}\,=\,\sum_{n=0}^{\infty} (\alpha')^n S_{D_p}^{(n)}\,,
\eeqa
and then information about $ S^{(n)}$ and $S_{D_p}^{(n)}$  can be found from the $\alpha'$-expansion of S-matrix elements \cite{Metsaev:1987zx,Gross:1986mw,Bachas:1999um,Corley:2001hg}  and from T-duality \cite{Kaloper:1997ux,Garousi:2009dj}.

The bulk effective action of bosonic string theory includes various couplings of the closed string tachyon $\tau$, graviton $G_{\mu\nu}$, dilaton $\phi$  and antisymmetric B-field.  Because of the tachyon, the perturbative bosonic string theory is unstable. In this paper we assume that the tachyon  freezes    at $\tau=0$. With this assumption, the leading $\alpha'$-order terms of the bulk effective action in the bosonic string theory  are given by the following couplings
\beqa
S^{(0)} =\frac{1}{2\kappa^2}\int d^{D}x e^{-2\phi}\sqrt{-G}\bigg[R+4(\prt\phi)^2-\frac{1}{12}H^2\bigg]\labell{bulk}\,.
\eeqa
The heterotic   and the superstring theories have the same couplings as well as some other couplings involving the other massless fields in their corresponding supermultiplets \cite{Becker:2007zj}.   The above action is    invariant under the standard Buscher rules for the  T-duality transformations (see \eg \cite{Hohm:2010jy}). 

The next-to-leading $\alpha'$-order terms of the bulk effective action  have been found in \cite{Metsaev:1987zx} from the corresponding sphere-level S-matrix  elements.  T-duality, however,  is not manifest in these couplings.   It has been shown in \cite{Meissner:1996sa,Kaloper:1997ux}   that  by  using proper field redefinitions, one can change the couplings at order $\alpha'$  into a  manifestly T-dual invariant form. However, the T-duality transformations are the standard Buscher rules plus some $\alpha'$-corrections \cite{Kaloper:1997ux}.  The T-dual invariant action is $S^{(0)}+\alpha' S^{(1)}$ where the action $S^{(0)}$ is given in \reef{bulk} and $S^{(1)}$ is 
\beqa
S^{(1)}&\!\!\!\!\!=\!\!\!\!\!&\frac{ \lambda_0}{2\kappa^2}\int d^{D}x e^{-2\phi}\sqrt{-G}\bigg[-R^2_{GB}+16(R^{\mu\nu}\!-\!\frac{1}{2}G^{\mu\nu}R)\prt_{\mu}\phi\prt_{\nu}\phi\!-\!16\nabla^2\phi(\prt\phi)^2+16(\prt\phi)^4\nonumber\\
&&+\frac{1}{2}(R_{\mu\nu\lambda\rho}H^{\mu\nu\alpha}H^{\lambda\rho}{}_{\alpha}\!-\!2R^{\mu\nu}H_{\mu\nu}^2+\frac{1}{3}RH^2)-2(\nabla^{\mu}\prt^{\nu}\phi H_{\mu\nu}^2-\frac{1}{3}\nabla^2\phi H^2)\!-\!\frac{2}{3}(\prt \phi)^2H^2\nonumber\\
&&-\frac{1}{24}H_{\mu\nu\lambda}H^{\nu}{}_{\rho\alpha}H^{\rho\sigma\lambda}H_{\sigma}{}^{\mu\alpha}+\frac{1}{8}H^2_{\mu\nu}H^{2\,\mu\nu}-\frac{1}{144}(H^2)^2\bigg]\labell{bulk2}\,,
\eeqa
where $\lambda_0=-\frac{1}{4}$ for the  bosonic string theory, $\lambda_0=-\frac{1}{8}$ for the heterotic theory and $\lambda_0=0$ for the superstring theory. In above action, $H^2_{\mu\nu}=H_{\mu\alpha\beta}H_{\nu}{}^{\alpha\beta}$ and  $R^2_{GB}=R^2_{\mu\nu\lambda\sigma}-4R^2_{\mu\nu}+R^2$ is the Gauss-Bonnet  combination of the curvature squared terms which does not change the graviton propagator in \reef{bulk}.   The T-duality invariance of the above action  is such that the  action $S^{(1)}$ itself is not fully invariant under the standard T-duality transformation.  It  produces some extra terms. The extra terms however are canceled with the transformation of the action $S^{(0)}$   under the $\alpha'$-corrected T-duality \cite{Kaloper:1997ux}.

Unlike the leading $\alpha'$-order action \reef{bulk}, the couplings in  \reef{bulk2} are not unique. One can use  field redefinitions of order $\alpha'$  on  the action \reef{bulk} to change the  couplings in the action \reef{bulk2} \cite{Metsaev:1986yb,Tseytlin:1986zz,Metsaev:1987zx}. 
 For example, using the field redefinition, 
\beqa
\phi\rightarrow \phi+\delta\phi\,, \qquad G_{\mu\nu}\rightarrow G_{\mu\nu}+\delta G_{\mu\nu}\,,
\eeqa
where 
\beqa
\delta\phi=\alpha'\lambda_0a_1H^2\,, \qquad \delta G_{\mu\nu}=\alpha'\lambda_0(b_1H_{\mu\nu}^2+b_2\prt_{\mu}\phi\prt_{\nu}\phi)\labell{fieldred}\,,
\eeqa
 one can produce the following couplings from \reef{bulk}:
\beqa
 &&\frac{\alpha' \lambda_0}{2\kappa^2}\int d^{D}x e^{-2\phi}\sqrt{-G}\bigg[-2a_1H^2 \Big(R+4\nabla^2\phi-4(\prt\phi)^2-\frac{1}{12}H^2\Big)\labell{bulk0}\\
&&-(b_1H_{\mu\nu}^2+b_2\prt_{\mu}\phi\prt_{\nu}\phi)\Big(R^{\mu\nu}\!+\!2\nabla^{\mu}\prt^{\nu}\phi\!-\!\frac{1}{4}H^{2\,\mu\nu}\!-\!\frac{1}{2}G^{\mu\nu}[R\!+\!4\nabla^2\phi\!-\!4(\prt\phi)^2\!-\!\frac{1}{12}H^2]\Big) \bigg]\,,\nonumber
\eeqa
where $a_1,\, b_1$ and $b_2$ are arbitrary constants. Adding the above terms into the action \reef{bulk2} for the following specific values:
\beqa
a_1=-\frac{1}{6}\,,\qquad b_1=-1\,,\qquad b_2=16\labell{val}\,,
\eeqa
  one   finds that the bulk action at order $\alpha'$  simplifies to 
\beqa
S^{(1)}&\!\!\!\!\!=\!\!\!\!\!&\frac{ \lambda_0}{2\kappa^2}\int d^{D}x e^{-2\phi}\sqrt{-G}\bigg[-R^2_{GB} +\frac{1}{2}R_{\mu\nu\lambda\rho}H^{\mu\nu\alpha}H^{\lambda\rho}{}_{\alpha}+32\nabla^2\phi(\prt\phi)^2-48(\prt\phi)^4 \labell{bulk3}\\
&&-\frac{2}{3}(\prt \phi)^2H^2+4\prt^{\mu}\phi\prt^{\nu}\phi H^2_{\mu\nu} -\frac{1}{24}H_{\mu\nu\lambda}H^{\nu}{}_{\rho\alpha}H^{\rho\sigma\lambda}H_{\sigma}{}^{\mu\alpha}-\frac{1}{8}H^2_{\mu\nu}H^{2\,\mu\nu}+\frac{1}{144}(H^2)^2\bigg]\nonumber\,,
\eeqa
where we have also used the integration by part to write $\nabla^{\mu}\prt^{\nu}\phi\prt_{\mu}\phi\prt_{\nu}\phi=-\frac{1}{2}\nabla^2\phi(\prt\phi)^2+(\prt\phi)^4$.    One may use a different field redefinition to write the above couplings in yet another form. However, in order 
  to keep the Gauss-Bonnet combination unchanged under the field redefinition, we are not allowed to use the field redefinitions   $\delta G_{\mu\nu}=\alpha'\lambda_0(b_3G_{\mu\nu}R+b_4 R_{\mu\nu})$ and $\delta\phi=\alpha'\lambda_0a_2 R$. Moreover, we are not allowed to have the field redefinitions $\delta B_{\mu\nu}\sim \nabla^{\alpha}H_{\alpha\mu\nu}$ or $\delta\phi\sim\nabla^2\phi$ because these change the B-field or  dilaton propagators, respectively. The couplings  \reef{bulk0} are not invariant under the T-duality. As a result, the above action is  not manifestly invariant under the T-duality.

The  effective action of D-brane  includes various  world volume couplings of open string tachyon $T$,  transverse scalar fields $\Phi^i$,   gauge field $A_a$, closed string tachyon, graviton, dilaton and B-field.  Duo to the presence of the open string tachyon, the D-branes in bosonic string theory are all unstable. We again assume that the open string tachyon freezes at $T=0$ and the closed string tachyon  at $\tau=0$. Then the leading $\alpha'$-order terms of the D-brane effective action are given by the Dirac-Born-Infeld
(DBI) action \cite{Leigh:1989jq,Bachas:1995kx}
\beqa
S^{(0)}_{D_p}=-T_p\int d^{p+1}x
e^{-\phi}\sqrt{-det(\tG_{ab}+\tB_{ab})}\,,\labell{dbi}
\eeqa
where $\tG_{ab}$ and $\tB_{ab}$ are pull-back of the bulk fields $G_{\mu\nu}$ and $B_{\mu\nu}$
onto the world-volume of the D-brane\footnote{Our index convention is such that  the Greek letters are used as space-time indices as usual.
The Latin letters $(a,b,c,...)$ denote the world-volume indices while $(i,j,k,...)$ denote 
the transverse or normal bundle indices.}, \eg 
\beqa
\tG_{ab}=\frac{\prt X^{\mu}}{\prt{\sigma^a}}\frac{\prt X^{\nu}}{\prt{\sigma^b}}G_{\mu\nu}\,.
\eeqa
 The gauge field is added to this action by the replacement $B\rightarrow B+2\pi\alpha'F$. The transverse scalar fields appear in the static gauge where $X^a=\sigma^a$ and $X^i=2\pi\alpha'\Phi^i$. The DBI action also describes the dynamics of  D-branes of superstring theory at low energy. This action is   covariant under   T-duality transformations \cite{Myers:1999ps}. 
  
 The $\alpha'$  corrections to the D-brane
action \reef{dbi} have been studied in  \cite{Corley:2001hg,Ardalan:2002qt}.  By requiring the consistency of the effective action with  $\alpha^{\prime}$-order
terms of the   disk-level scattering amplitude of two gravitons, the following gravity couplings have  been found in  \cite{Corley:2001hg}:
\beqa
S_{D_p}^{(1)}=
-\frac{ T_p}{2}\int d^{p+1}xe^{-\phi}
\sqrt{-\tG }\bigg[\tR+\bot_{\alpha\beta}(\Omega^\alpha{}_{ a}{}^a
\Omega^\beta{}_{ b}{}^b-{\Omega^{\alpha}}_{ab}
\Omega^{\beta\,ab}) +\dots\bigg]\labell{dbi1}\,,
\end{eqnarray}
where $\tR$ is the scalar curvature made out of the pull-back  metric  $\tG_{ab}$ and 
 $\Omega$ is the second fundamental form 
 \beqa
 {\Omega^{\alpha}}_{ab}=\frac{ \prt^2 X^\alpha}{\prt\sigma^a\prt\sigma^b}+\frac{\prt X^{\mu}}{\prt{\sigma^a}}\frac{\prt X^{\nu}}{\prt{\sigma^b}}\Gamma^\alpha_{\mu\nu}\,.
 \eeqa
The operator   $\bot^{\mu\nu}$ is the  projection operator to the transverse space, \ie
\beqa
\bot^{\mu\nu}=G^{\mu\nu}-\tG^{\mu\nu}\,,\qquad\tG^{\mu\nu}=\frac{\prt X^{\mu}}{\prt{\sigma^a}}\frac{\prt X^{\nu}}{\prt{\sigma^b}}\tG^{ab}\,,
\eeqa
where $\tG^{\mu\nu}$  which   projects operators to the world volume space, is the first fundamental form, and the indices of the operator  $ \bot^{\mu\nu}$ in \reef{dbi1} are lowered by the spacetime metric\footnote{The relation between the second fundamental form $K^i_{ab}$ that has been used in \cite{Corley:2001hg} and $\Omega^{\mu}_{ab}$ is 
$
K^i_{ab}=\Omega^{\mu}_{ab}n_{\mu}^i
$
where $n_{\mu}^i$ for $i=p+1,\cdots, D$,  are the orthonormal basis of the transverse space, \ie $\bot_{\mu\nu}=\sum_{i=p+1}^D n_{\mu}^in_{\nu}^i$.}.

 The dots in  \reef{dbi1} refer to the dilaton and B-field couplings. The coefficients of these  couplings that have been found in \cite{Corley:2001hg}, however,  depend on  the dimension of the D$_p$-brane. We refer the interested readers to \cite{Corley:2001hg} for the explicit form of these couplings. Under the T-duality transformations, the graviton transforms into B-field, so in a T-duality invariant action the coefficients of the B-field and dilaton, like the coefficients of the gravity couplings in \reef{dbi1}, must be independent of $p$. As a result, the above action is not manifestly invariant under the T-duality.
 
In this paper, we are going to find an action which is   manifestly invariant under  the T-duality. We have found that it is impossible to find appropriate dilaton and B-field couplings which make the gravity couplings in \reef{dbi1}   to be invariant under the standard T-duality transformations. However, when  we use the $\alpha'$-corrected T-duality transformations \cite{Kaloper:1997ux}, we will find the following T-duality  invariant couplings:
\beqa
S_{D_p}^{(1)} \!\!\!&=&\!\!\!
-\frac{ T_p}{2}\int d^{p+1}xe^{-\phi}
\sqrt{-\tG }\bigg[\tR+\bot_{\mu\nu}(\Omega^\mu{}_{ a}{}^a
\Omega^\nu{}_{ b}{}^b-{\Omega^{\mu}}_{ab}
\Omega^{\nu\,ab})  +2\bot_{\mu\nu} 
\Omega^{\mu}{}_a{}^a\prt^{\nu}\phi   + \prt_{\mu}\phi\prt^{\mu}\phi \nonumber\\
&&-\frac{1}{8}\tilde{H}^2- \frac{1}{8} \bot^{\mu\nu}
H^2_{\mu\nu}   +\frac{1}{8} \bot^{\alpha\beta} \bot^{\mu\nu}
H_{\alpha\mu\lambda}H_{\beta\nu}{}^{\lambda} + \frac{1}{24} \bot^{\alpha\beta} \bot^{\mu\nu}\bot^{\lambda\sigma} H_{\alpha\mu\lambda}H_{\beta\nu\sigma}  \bigg]\labell{dbi2}\,,
\end{eqnarray}
where $\tilde{H}^2=H_{abc}H^{abc}$.  The above  D-brane action is consistent with  the bulk action \reef{bulk2}, \ie they  both are  manifestly invariant under the T-duality.  Similar calculations have been done in \cite{Garousi:2009dj,Garousi:2011fc} to find the T-dual completion of the gravity couplings on the world volume of D-branes in the superstring theory.

The actions \reef{bulk2} and \reef{dbi2} may be used to study  various physical phenomena like the scattering of external particles from the D-branes in the bosonic string theory. 
However, if one prefers to use  the simplified bulk action \reef{bulk3} in which the field redefinition \reef{fieldred}  has been used, then the same field redefinition must  be applied  on the DBI action to modify the brane action \reef{dbi2}.  
The field redefinition \reef{fieldred} produces the following couplings from the DBI action:
\beqa
-\frac{\alpha'T_p}{2}\int d^{p+1}xe^{-\phi}
\sqrt{-\tG } \bigg[-2a_1H^2+b_1\tG^{ab}H^2_{ab}+b_2\prt^a\phi\prt_a\phi\bigg]\lambda_0\,.
\eeqa
Adding the above couplings for the specific values \reef{val} to the action \reef{dbi2}, one finds
\beqa
S_{D_p}^{(1)} \!\!\!&=&\!\!\!
-\frac{ T_p}{2}\int d^{p+1}xe^{-\phi}
\sqrt{-\tG }\bigg[\tR+\bot_{\mu\nu}(\Omega^\mu{}_{ a}{}^a
\Omega^\nu{}_{ b}{}^b-{\Omega^{\mu}}_{ab}
\Omega^{\nu\,ab}) \labell{dbi3}-3\prt_a\phi\prt^a\phi \nonumber\\
&&\!\!\!+\bot_{\mu\nu}\prt^{\mu}\phi\prt^{\nu}\phi+2\bot_{\mu\nu} 
\Omega^{\mu}{}_a{}^a\prt^{\nu}\phi  +\frac{1}{24}\tilde{H}^2 + \frac{1}{8} \bot^{\mu\nu}
H^2_{ \mu\nu} \nonumber\\
&&\!\!\!
-\frac{3}{8} \bot^{\alpha\beta} \bot^{\mu\nu}
H_{\alpha\mu\lambda}H_{\beta\nu}{}^{\lambda} + \frac{5}{24} \bot^{\alpha\beta} \bot^{\mu\nu}\bot^{\lambda\sigma} H_{\alpha\mu\lambda}H_{\beta\nu\sigma} \bigg]\,,
\end{eqnarray}
which is the D-brane action corresponding to the bulk action \reef{bulk3}.  Unlike the D-brane action \reef{dbi2}, the above action is not manifestly invariant under the T-duality transformations. To compare the D-brane actions \reef{dbi2} or \reef{dbi3} with the string theory S-matrix elements, we have to transform them to the Einstein frame and use some other field redefinitions. Then we compare them with the corresponding disk-level S-matrix elements.

An outline of the paper is as follows: In section 2 we give a brief review of the standard T-duality transformation along with  its $\alpha'$ corrections \cite{Kaloper:1997ux}, and review the  prescription    given in \cite{Garousi:2009dj}
for finding the T-dual completion of a gravity coupling in the effective action of D-branes.  In section 3, using this method we find the appropriate $B$ field
and dilaton couplings which make the graviton couplings in  \reef{dbi1} to be invariant under the T-duality. In section 4, in order to compare the  action \reef{dbi3} with the  string theory S-matrix elements,  we transform the   actions \reef{bulk3} and \reef{dbi3} to the  Einstein frame. We find perfect agreement between the B-field couplings that we have found  and the B-field couplings that have been found in \cite{Corley:2001hg} from S-matrix calculations. The dilaton couplings, however, are not exactly the couplings that have been found in \cite{Corley:2001hg}. In Appendix A, we   reexamine the extraction of the dilaton couplings from the S-matrix element of two gravitons and find an exact agreement with the couplings that we have found from T-duality.

\section{T-duality}

The full set of nonlinear T-duality transformations for massless  fields have been found in \cite{TB,Meessen:1998qm,Bergshoeff:1995as,Bergshoeff:1996ui}.   When  the T-duality transformation acts along the Killing coordinate $y$,  the transformations are
\beqa
&&\,\,e^{2\tphi}=\frac{e^{2\phi}}{G_{yy}}\,,\quad\quad\,
\tG_{yy}=\frac{1}{G_{yy}}\,,\nonumber\\
&&\tG_{\mu y}=\frac{B_{\mu y}}{G_{yy}}\,,\qquad
\tG_{\mu\nu}=G_{\mu\nu}-\frac{G_{\mu y}G_{\nu y}-B_{\mu y}B_{\nu y}}{G_{yy}}\,,\nonumber\\
&&\tB_{\mu y}=\frac{G_{\mu y}}{G_{yy}}\,,\qquad
\tB_{\mu\nu}=B_{\mu\nu}-\frac{B_{\mu y}G_{\nu y}-G_{\mu y}B_{\nu y}}{G_{yy}} \,,\labell{Cy1}
\eeqa
where $\mu,\nu\ne y$. In above transformation the metric is given in the string frame. If $y$ is identified on a circle of radius $\rho$, \ie $y\sim y+2\pi \rho$, then after T-duality the radius becomes $\tilde{\rho}=\alpha'/\rho$. The string coupling is also transformed as $\tilde{g}=g\sqrt{\alpha'}/\rho$.

It is known that the T-duality transformations in the  superstring theory do not receive  $\alpha'$ corrections, however, they receive such corrections in the  heterotic and  bosonic string theories \cite{Kaloper:1997ux}.  That is, the T-duality operator has an $\alpha'$ expansion
\beqa
T&=&\sum_{n=0}^{\infty}(\alpha')^nT^{(n)}\,,
\eeqa
where $T^{(n)}$ for $n>0$ are all zero in the superstring theory and are non-zero in other cases. In all theories $T^{(0)}$ is given by the Buscher rules \reef{Cy1}. The invariance of the effective actions at  order $(\alpha')^0$  then means that
\beqa
S^{(0)}&\stackrel{T^{(0)}}{\longrightarrow}&S^{(0)}\,.
\eeqa
At order $\alpha'$, the action has two terms, \ie $S=S^{(0)}+\alpha' S^{(1)}$. The invariance then means
\beqa
S^{(1)}&\stackrel{T^{(0)}}{\longrightarrow}&S^{(1)}+\delta S\,,\nonumber\\
S^{(0)}&\stackrel{T^{(1)}}{\longrightarrow}&-\delta S\,.
\eeqa
At order $(\alpha')^2$, the action has three terms, \ie $S=S^{(0)}+\alpha' S^{(1)}+(\alpha')^2 S^{(2)}$ and again the invariance means that
\beqa
S^{(2)}&\stackrel{T^{(0)}}{\longrightarrow}&S^{(2)}+\delta S_1+\delta S_2\,,\nonumber\\
S^{(1)}&\stackrel{T^{(1)}}{\longrightarrow}&-\delta S_1\,,\nonumber\\
S^{(0)}&\stackrel{T^{(2)}}{\longrightarrow}&-\delta S_2\,.
\eeqa
Similarly for the action at higher orders of $\alpha'$.

To study  the $\alpha'$ corrections to the T-duality transformations, it is convenient to   introduce the following new fields:  
\beqa
&&g_{\mu\nu}\equiv G_{\mu\nu}-G^{yy}G_{\mu y}G_{\nu y}\,,\qquad b_{\mu\nu}\equiv B_{\mu\nu}-\frac{1}{2}G^{yy}(B_{\mu y}G_{\nu y}- B_{\nu y}G_{\mu y})\,,\nonumber\\
&&\,\,\,\,\bar{\phi}\equiv\phi-\frac{1}{4}\ln G_{yy}\,,\quad
V_{\mu}\equiv\sqrt{G^{yy}}G_{\mu y}\,,\quad W_{\mu}\equiv\sqrt{G^{yy}}B_{\mu y}\,,\quad \sigma\equiv\frac{1}{2}\ln G_{yy}\,.
\eeqa
In terms of these fields, the T-duality operator at leading $\alpha'$-order $T^{(0)}$  simplifies as
\beqa
\sigma\stackrel{T^{(0)}}\longrightarrow -\sigma\,,\qquad V_{\mu}\stackrel{T^{(0)}}\longrightarrow W_{\mu}\,,\qquad W_{\mu}\stackrel{T^{(0)}}\longrightarrow V_{\mu}\labell{Cy}\,.
\eeqa
The fields $g_{\mu\nu},\ b_{\mu\nu}$ and $\bar{\phi}$ remain invariant under the T-duality. Up to the order of $\alpha'$, the T-duality operator has been found in \cite{Kaloper:1997ux} to be 
\beqa
\sigma&\stackrel{T}\longrightarrow &-\sigma-\frac{\alpha'\lambda_0}{2}\bigg[8(\nabla\sigma)^2+e^{2\sigma}V_{\mu\nu}V^{\mu\nu}+e^{-2\sigma}W_{\mu\nu}
W^{\mu\nu}\bigg]\,,\nonumber\\
V_{\mu}&\stackrel{T}\longrightarrow &W_{\mu}-\alpha'\lambda_0\,\bigg[4W_{\mu\nu}\nabla^{\nu}\sigma+e^{2\sigma}H_{\mu\nu\lambda}V^{\nu\lambda}\bigg]\,,\nonumber\\
W_{\mu}&\stackrel{T}\longrightarrow &V_{\mu}-\,\alpha'\lambda_0\,\,\bigg[4V_{\mu\nu}\nabla^{\nu}\sigma-e^{-2\sigma}H_{\mu\nu\lambda}W^{\nu\lambda}\bigg]\,, \nonumber\\
H_{\mu\nu\lambda}&\stackrel{T}\longrightarrow &H_{\mu\nu\lambda}-12\alpha'\lambda_0\bigg[\nabla_{[\mu}(W_{\nu}{}^{\rho}V_{\lambda]\rho})+V_{[\mu\nu}W_{\lambda]\rho}\nabla^2\sigma
+W_{[\mu\nu}V_{\lambda]\rho}\nabla^2\sigma\,, \nonumber\\
&&\qquad\,\,\,+\frac{1}{4}e^{2\sigma}V^{\rho\chi}V_{[\mu\nu}H_{\lambda]\rho\chi}-\frac{1}{4}e^{-2\sigma}W^{\rho\chi}W_{[\mu\nu}H_{\lambda]\rho\chi}\bigg]\,.\labell{corrT}
\eeqa
The metric $g_{\mu\nu}$ and $\bar{\phi}$ remain invariant. In above transformations, $H$ is the field strength of the b-field, \ie $H_{\mu\nu\lambda}=\prt_{\mu}b_{\nu\lambda}+\prt_{\lambda}b_{\mu\nu }+\prt_{\nu}b_{\lambda\mu}$, $V_{\mu\nu}$ is the field strength of $V_{\mu}$, \ie $V_{\mu\nu}=\prt_{\mu}V_{\nu}-\prt_{\nu}V_{\mu}$ and $W_{\mu\nu}$ is the field strength of $W_{\mu}$, \ie $W_{\mu\nu}=\prt_{\mu}W_{\nu}-\prt_{\nu}W_{\mu}$.

One may assume that the metric is a small perturbation around the flat space, \ie $G_{\alpha\beta}=\eta_{\alpha\beta}+h_{\alpha\beta}$, and also assume that the dilaton and B-field  are small perturbations. Then one can find  a  perturbative expansion for the T-duality transformation of the graviton, dilaton and B-field.

A perturbative  method for  finding   the T-duality invariant world volume couplings is given in \cite{Garousi:2009dj}.   Let us review this method here.  
A coupling in general has world volume and normal bundle  indices. Suppose we are implementing T-duality along a world volume direction $y$ of a D$_p$-brane.   We first separate the  world-volume indices  along and orthogonal to $y$,     and then apply the  T-duality transformations \reef{corrT}.  The  orthogonal indices  are  the complete world-volume indices   of the  T-dual D$_{p-1}$-brane. However,  $y$  in the T-dual theory  which is a normal bundle index, is not  complete. 
On the other hand, the normal bundle indices of  the original theory  are not complete in the T-dual D$_{p-1}$-brane. They do not include the $y$ index.  In a T-duality invariant theory, the index  $y$   must be combined  with the incomplete normal bundle indices  to make them complete.  This last step can be done by rewriting an incomplete normal bundle index  as a complete normal bundle  index minus the $y$-index. Then all the couplings which have the $y$-index must be canceled.  If the world volume couplings with the $y$-index are not canceled,  one should then add  new couplings  to the original theory  to be able to cancel them.

\section{T-dual completion of  the gravity couplings}

In this section we are going to apply the perturbative method  outlined in the previous section to extend the   brane action \reef{dbi1} to be  invariant under the T-duality transformations \reef{corrT}.  Let us first review how the gravity couplings \reef{dbi1} have been found in \cite{Corley:2001hg}. The authors of \cite{Corley:2001hg}  consider various gravity couplings at order $\alpha'$  with arbitrary coefficients, \ie
 \beq
 \frac{\alpha'T_p}{2}\!\!\!\int\!\! d^{p+1}\!xe^{-\phi}
\sqrt{\!-\tG }\bigg[\beta_0\tR \!+\!\beta_1 \bot_{\mu\nu}{\Omega^\mu}_{ a}{}^a
{\Omega^\nu}_{b}{}^b\!+\!\beta_2\bot_{\mu\nu}{\Omega^\mu }_{ab}
{\Omega^\nu}^{ab}\!+\!\beta_3\bot^{\mu\nu}R_{\mu\nu}\!+\!\beta_4\bot^{\mu\nu}\!\bot^{\alpha\beta}\!R_{\mu\alpha\nu\beta}\bigg].\nonumber
\eeq
There is another gravity coupling at this order which is given by the bulk scalar curvature $R$ evaluated on the D-brane.  However, it is not independent according to  the Gauss identity (see \eg \cite{Carter:1997pb})
\beqa
\tR-R+2\bot^{\mu\nu}R_{\mu\nu}-\bot^{\mu\nu} \bot^{\alpha\beta}R_{\mu\alpha\nu\beta}-\bot_{\mu\nu}({\Omega^\mu}_{ a}{}^a
{\Omega^\nu}_{ b}{}^b-{\Omega^\mu }_{ab}
{\Omega^\nu }^{ab})=0\,.\labell{Gauss}
\eeqa
Using the fact that the $\alpha'$ terms should not change the propagator of the transverse scalars, the relation $\beta_{1}=-\beta_2$ has been found. Then using these D-brane couplings and the Gauss-Bonnet couplings in the bulk action,  the massless poles and the contact terms of the   scattering  amplitude of two gravitons from D-brane have been calculated. By equating these terms with  the corresponding terms in the disk-level S-matrix element of two gravitons, one finds the gravity couplings in \reef{dbi1} uniquely.

 Now to extend the gravity couplings \reef{dbi1} to be invariant under the T-duality, one has to again consider all the B-field and dilaton couplings at  order $\alpha'$ with arbitrary coefficients, \ie
\beqa 
 &&\frac{\alpha'T_p}{2}\int d^{p+1}xe^{-\phi}
\sqrt{-\tG }\bigg[ \alpha_1H_{abc}H^{abc}+\alpha_2  \bot^{\alpha\beta}
H_{\alpha \mu\nu}H_{\beta}{}^{\mu\nu}+\alpha_3 \bot^{\alpha\beta} \bot^{\mu\nu}  H_{\alpha\mu\lambda}H_{\beta\nu }{}^{\lambda}\nonumber\\
&&\qquad\qquad\qquad\qquad\qquad+\alpha_4 \bot^{\alpha\beta} \bot^{\mu\nu}\bot^{\lambda\sigma} H_{\alpha\mu\lambda}H_{\beta\nu\sigma}+\alpha_5B_{ab}\nabla_{\mu}H^{\mu a b}+\sigma_1 G^{\mu\nu}\nabla_{\mu}\prt_{\nu}\phi\nonumber\\
&&\qquad\qquad\qquad\qquad\qquad+\sigma_2\bot^{\mu\nu}\nabla_{\mu}\prt_{\nu}\phi+\sigma_3\tG^{ab}\prt_a\phi\prt_b\phi+\sigma_4G^{\mu\nu}\prt_{\mu}\phi\prt_{\nu}\phi\bigg]\,.\labell{add}
\end{eqnarray}
Note that the  coupling $H^2$ is not independent of the other B-field couplings that we have considered, \ie
\beqa
H^2= H_{abc}H^{abc}+3 \bot^{\alpha\beta}
H_{\alpha \mu\nu}H_{\beta}{}^{\mu\nu}-3 \bot^{\alpha\beta} \bot^{\mu\nu}  H_{\alpha\mu\lambda}H_{\beta\nu }{}^{\lambda} +  \bot^{\alpha\beta} \bot^{\mu\nu}\bot^{\lambda\sigma} H_{\alpha\mu\lambda}H_{\beta\nu\sigma}\,.\labell{Hiden}
\eeqa
Neither the dilaton coupling $\bot^{\mu\nu}\prt_{\mu}\phi\prt_{\nu}\phi$ nor the coupling $ \bot^{\mu\nu}\Omega^{\mu}{}_a{}^a\prt^{\nu}\phi$ are independent, \ie 
\beqa
 \bot^{\mu\nu}\prt_{\mu}\phi\prt_{\nu}\phi &=&G^{\mu\nu}\prt_{\mu}\phi\prt_{\nu}\phi-
\tG^{ab}\prt_a\phi\prt_b\phi\,,\nonumber\\
\tG^{\mu\nu}\nabla_{\mu}\prt_{\nu}\phi &=&\tilde{\nabla}^a\prt_a\phi-\bot_{\mu\nu}\Omega^{\mu}{}_a{}^a\prt^{\nu}\phi\,.\labell{phiiden}
\eeqa
 Moreover, up to a total derivative term, one has the identity $\tilde{\nabla}^a\prt_a\phi=\prt^a\phi\prt_a\phi$ in the string frame action. So the  terms in \reef{add} are all independent B-field and dilaton couplings at order $\alpha'$.

After adding the couplings \reef{add} to the gravity couplings \reef{dbi1}, we use the perturbation $G_{\mu\nu}=\eta_{\mu\nu}+h_{\mu\nu}$ and keep the terms with linear- and quadratic-order fields. Following the method outlined in the previous section, we separate the world volume indices along and orthogonal to the Killing index $y$. Then we use the   T-duality transformations \reef{Cy}. In the T-dual theory we also complete the transverse indices. Doing all these steps, we have found that it is impossible to cancel the couplings which have the $y$-index.   This confirms the observation made in \cite{Kaloper:1997ux} that the T-duality in the bosonic and the heterotic string theories must receive $\alpha'$ corrections \reef{corrT}.
 
 The $\alpha'$-corrected T-duality transformations \reef{corrT} must be applied on the DBI action \reef{dbi}. Since we keep the perturbative fields up to the quadratic order, we have to take into account only the $\alpha'$ corrections in the first line of \reef{corrT}. The $\alpha'$ terms in \reef{corrT} start at quadratic order terms, so we have to consider the linear order terms in the DBI action which are
 \beq
 S_{D_p}^{(0)}=-T_p\int d^{p+1}x\,(1-\phi+\frac{1}{2}\eta^{ab}h_{ab}+\cdots)\,,
 \eeq
 where dots refer to the higher-order terms. Now we have to separate the world volume indices along and orthogonal to $y$, \ie
 \beq
 S_{D_p}^{(0)}=-T_p\int d^{p+1}x\,(1-\phi+\frac{1}{2}h_{yy}+\frac{1}{2}\eta^{\ta\tb}h_{\ta\tb}+\cdots)\,,
 \eeq
where the world volume  indices $\ta,\, \tb$ are orthogonal to $y$.  Under the T-duality \reef{corrT}, $-\phi+\frac{1}{4}h_{yy}$ is invariant, and the $\alpha'$-corrected T-duality of $h_{\ta\tb}$ has no quadratic order term. So the T-duality of above action is given by the T-duality transformation of $\frac{1}{4}h_{yy}$ which is
 \beqa
 \frac{1}{4}h_{yy}\!\stackrel{T}{\longrightarrow}\! -\frac{1}{4}h_{yy}+\frac{\alpha'}{8}\bigg[\prt_{\mu}h_{yy}\prt_{\mu}h_{yy}+\prt_{\mu}h_{\nu y}(\prt_{\mu}h_{\nu y}-\prt_{\nu}h_{\mu y})+\prt_{\mu}B_{\nu y}(\prt_{\mu}B_{\nu y}-\prt_{\nu}B_{\mu y})\bigg].
 \eeqa
Taking  into account the above $y$-dependent terms at order $\alpha'$  and ignoring some total derivative terms, one is able to find the T-dual completion of the gravity couplings.    With the assistance of the computer algebra system, ``Cadabra'', \cite{Peeters:2006kp,Peeters:2007wn}, we have found the result  in \reef{dbi2}.  


\section{Couplings in  the Einstein frame}
 
To compare the actions \reef{dbi2} or \reef{dbi3}  with the S-matrix elements in string theory, we have to transform them to the   Einstein frame  $G^s_{\mu\nu}=e^{\gamma \phi }G^E_{\mu\nu}$ where $\gamma=4/(D-2)$.  Since the S-matrix elements are independent of field redefinitions, we    use the simplified actions  in which the field redefinitions have been used. We first transform the bulk actions to the Einstein frame. 

For those terms  which have no  derivative of the metric, the transformation  gives only an overall dilaton factor.  In other cases, there are some extra terms involving the derivative of the dilaton, \eg the transformation of scalar curvature is 
\beqa
R&\Longrightarrow&e^{-\gamma\phi}\bigg[R-\gamma(D-1)\nabla^2\phi-\frac{\gamma^2}{4} (D-1)(D-2)
(\prt\phi)^2\bigg]\,,
\eeqa
where on the right hand side the metric is in the Einstein frame. This transforms the string frame action \reef{bulk} to the following action in the Einstein frame:
\beqa
S^{(0)}=\frac{1}{2\kappa^2}\int d^{D}x  \sqrt{-G}\bigg[R-\gamma(\prt\phi)^2-\frac{1}{12}e^{-2\gamma\phi}H^2\bigg]\,,\labell{bulkE}
\eeqa
where the total derivative term has been dropped.

	To transform the $\alpha'$-order terms in \reef{bulk3}, one needs the following relations between the two frames:
\beqa
\nabla_{\mu}\prt_{\nu}\phi &\Longrightarrow&\nabla_{\mu}\prt_{\nu}\phi -\frac{\gamma}{2}\Big(2\prt_{\mu}\phi\prt_{\nu}\phi-G_{\mu\nu}\prt_{\alpha}\phi\prt^{\alpha}\phi\Big)\,,\labell{rel}\\
R_{\mu\nu\alpha\beta}&\Longrightarrow&e^{\gamma\phi}R_{\mu\nu\alpha\beta}+2\gamma e^{\gamma\phi}\bigg[G_{[\mu[\beta}\nabla_{\nu]}\prt_{\alpha]}\phi+\frac{\gamma}{2}G_{[\mu[\alpha}\prt_{\nu]}\phi\prt_{\beta]}\phi+\frac{\gamma}{4}G_{[\mu[\beta}G_{\nu]\alpha]}\prt_{\lambda}\phi\prt^{\lambda}\phi\bigg]\,.\nonumber
\eeqa
 Using these transformations, one   finds the transformation of various terms   in the action \reef{bulk3}, \ie
\beqa
R_{\mu\nu\lambda\rho}H^{\mu\nu\alpha}H^{\lambda\rho}{}_{\alpha}\!\!\!\!&\Longrightarrow&\!\!\!\! e^{-4\gamma\phi}\bigg[R_{\mu\nu\lambda\rho}H^{\mu\nu\alpha}H^{\lambda\rho}{}_{\alpha}\!-\!2\gamma\nabla^{\mu}\prt^{\nu}\phi H^2_{\mu\nu}\!+\!\gamma^2\prt^{\mu}\phi\prt^{\nu}\phi H^2_{\mu\nu}\!-\!\frac{1}{2}\gamma^2(\prt\phi)^2H^2\bigg],\nonumber\\
\nabla^2\phi \!\!\!\!&\Longrightarrow&\!\!\!\! e^{-\gamma\phi}\,\,\bigg[\nabla^2\phi +2(\prt\phi)^2\bigg]\,,\nonumber\\
R^2_{GB}\!\!\!\!&\Longrightarrow&\!\!\!\! e^{-2\gamma\phi}\bigg[R^2_{GB}+4\gamma(D-3)\Big(R_{\mu\nu}-\frac{1}{2} G_{\mu\nu}R\Big)\nabla^{\mu}\prt^{\nu}\phi\nonumber\\
&&\qquad\quad
+4\gamma (D-3)\Big((\nabla^2\phi)^2-\nabla_{\mu}\prt_{\nu}\phi\nabla^{\mu}\prt^{\nu}\phi\Big)\nonumber\\
&&\qquad\quad-\frac{\gamma^2}{2}(D-3)\Big((D-4)(\prt\phi)^2R+4\prt_{\mu}\phi\prt_{\nu}\phi R^{\mu\nu}\Big)\nonumber\\
&&\qquad\quad+2\gamma^2 (D-3)\Big((D-3)\nabla^2\phi(\prt\phi)^2+2\nabla_{\mu}\prt_{\nu}\phi\prt^{\mu}\phi\prt^{\nu}\phi\Big)\nonumber\\
&&\qquad\quad+\frac{\gamma^3}{4}(D-1)(D-3)(D-4)(\prt\phi)^4\bigg]\,.
\eeqa
In the string frame, the Gauss-Bonnet term does not change the graviton propagator, so one expects that it does not change the graviton and dilaton propagators when transforming it to the  Einstein frame. In fact one can easily show that   the  quadratic order terms in the Gauss-Bonnet  can be written as  cubic order terms. To this end, consider the following identities:
\beqa
(R_{\mu\nu}-\frac{1}{2} G_{\mu\nu}R)\nabla^{\mu}\prt^{\nu}\phi&=&\nabla^{\mu}[(R_{\mu\nu}-\frac{1}{2} G_{\mu\nu}R) \prt^{\nu}\phi]\,,\nonumber\\
(\nabla^2\phi)^2-\nabla_{\mu}\prt_{\nu}\phi\nabla^{\mu}\prt^{\nu}\phi&=&\nabla_{\mu}[\prt^{\mu}\phi\nabla^2\phi]-\nabla_{\mu}[\prt_{\nu}\phi\nabla^{\mu}\prt^{\nu}\phi]+\prt_{\mu}\phi\prt_{\nu}\phi R^{\mu\nu}\,,
\eeqa
where in the first equation we have used the Bianchi identity $\nabla^{\mu}R_{\mu\nu}-\frac{1}{2}\nabla_{\nu}R=0$, and in the second equation we have used the (non)commutative property of the covariant derivative, \ie  $[\nabla_{\mu},\nabla_{\nu}]A^{\alpha}=R^{\alpha}{}_{\beta\mu\nu}A^{\beta}$ and $[\nabla_{\mu},\nabla_{\nu}]\phi=0$.  The above total derivative terms, however, can not be dropped because of the overall dilaton  factor  $e^{-\gamma\phi}$ in the Einstein frame. Using the integration by part, one can write the Gauss-Bonnet term as 
\beqa
e^{-2\phi}\sqrt{-G}R^2_{GB}&\Longrightarrow&e^{-\gamma\phi}\sqrt{-G}\bigg[R^2_{GB}  
+ \gamma^2 D(D-3) \prt_{\mu}\phi\prt_{\nu}\phi (R^{\mu\nu}-\frac{1}{2}G^{\mu\nu}R)\nonumber\\
&&\qquad\qquad\quad+2\gamma^2 (D-1)(D-3)\nabla^2\phi(\prt\phi)^2 \nonumber\\
&&\qquad\qquad\quad+\frac{\gamma^3}{4}(D-1)(D-3)(D-4)(\prt\phi)^4\bigg]\,.
\eeqa
It is interesting to note that the dilaton terms vanish in three dimensions. This is consistent with the fact that the Gauss-Bonnet term is zero in three dimensions.

Using the above transformations, one finds that the action \reef{bulk3} transforms to the following action in   the Einstein frame:
\beqa
S^{(1)}&\!\!\!\!\!=\!\!\!\!\!&\frac{ \lambda_0}{2\kappa^2}\int d^{D}x \,e^{-\gamma\phi}\sqrt{-G}\bigg[-R^2_{GB} +e^{-2\gamma\phi}\Big(\frac{1}{2}R_{\mu\nu\lambda\rho}H^{\mu\nu\alpha}H^{\lambda\rho}{}_{\alpha}-\gamma\nabla^{\mu}\prt^{\nu}\phi H^2_{\mu\nu}\Big) 
\labell{bulkE4}\\
&& -\gamma^2 D(D-3) \prt_{\mu}\phi\prt_{\nu}\phi \Big(R^{\mu\nu}-\frac{1}{2}G^{\mu\nu}R\Big)\!-\!\Big(2\gamma^2(D-1) (D-3)-32\Big)\nabla^2\phi(\prt\phi)^2\!+\!\cdots \!\!\bigg],\nonumber
\eeqa
where  dots represent the quartic order terms. The above action can be simplified by using field redefinition in the Einstein frame. Since we are interested in actions which can be compared with the S-matrix elements, we are not going to use the field redefinitions which change the Gauss-Bonnet combination of the curvature squared. Moreover, the Riemann curvature in the second term above can not be changed under the field redefinition. All other cubic terms in the above action  can  be  converted to some quartic order terms  by an appropriate  field redefinition.  

The dilaton and the metric variations of the  action \reef{bulkE}  are  
\beqa
 &&\frac{1}{2\kappa^2}\int d^{D}x \sqrt{-G}\bigg[\delta\phi\Big(2\gamma\nabla^2\phi+\frac{\gamma}{6}e^{-2\gamma\phi}H^2\Big)-\delta G_{\mu\nu}\Big(R^{\mu\nu}-\gamma\prt^{\mu}\phi\prt^{\nu}\phi\nonumber\\
&&\qquad\qquad \qquad\quad  -\frac{1}{4} e^{-2\gamma\phi}H^{2\,\mu\nu}-\frac{1}{2}G^{\mu\nu}[R-\gamma(\prt\phi)^2 -\frac{1}{12} e^{-2\gamma\phi}H^2]\Big) \bigg] \,.\nonumber
\eeqa
Using the following field redefinitions:
\beqa
\delta\phi&=&\alpha'\lambda_0 e^{-\gamma\phi}\bigg(\gamma (\prt\phi)^2+\frac{1}{12}e^{-2\gamma\phi}H^2\bigg)\,, \nonumber\\
\delta G_{\mu\nu}&=&- \alpha'\lambda_0e^{-\gamma\phi}\bigg(\gamma^2 [D(D-3)+4]
\prt_{\mu}\phi\prt_{\nu}\phi-4 \gamma\nabla_{\mu}\prt_{\nu}\phi \bigg)\,, \labell{red2}
\eeqa
one finds that  the Einstein frame action \reef{bulkE4} converts to the following standard form of the bulk action at order $\alpha'$: 
\beqa
S^{(1)}\!\!\!\!& = &\!\!\!\!\frac{ \lambda_0}{2\kappa^2}\!\int\! d^{D}x \,e^{-\gamma\phi}\sqrt{-G}\bigg[-R^2_{GB}  
 +\frac{1}{2}e^{-2\gamma\phi}R_{\mu\nu\lambda\rho}H^{\mu\nu\alpha}H^{\lambda\rho}{}_{\alpha}-\frac{\gamma^2(D-4)}{(D-2)}(\prt\phi)^4 \nonumber\\
&&\qquad\qquad\qquad\qquad\quad-e^{-2\gamma\phi}\Big(\gamma\prt^{\mu}\phi\prt^{\nu}\phi H^2_{\mu\nu}-\frac{\gamma}{6}(\prt\phi)^2H^2\Big)-e^{-4\gamma\phi}\Big(\frac{1}{8}H^2_{\mu\nu}H^{2\,\mu\nu}\nonumber\\
&&\qquad\qquad\qquad\qquad\quad +\frac{1}{24}H_{\mu\nu\lambda}H^{\nu}{}_{\rho\alpha}H^{\rho\sigma\lambda}H_{\sigma}{}^{\mu\alpha}-\frac{(D+6)}{144(D-2)}(H^2)^2\Big)\bigg]\,.\labell{bulkE3}
\eeqa
  These terms   are exactly the couplings that have been found in \cite{Metsaev:1987zx} from the corresponding sphere-level S-matrix elements.

We now transform the brane actions to the Einstein frame. The string frame DBI action \reef{dbi} transforms to the following action:
\beqa
S_{D_p}^{(0)}=-T_p\int d^{p+1}x
e^{-\phi\big[1-\gamma(p+1)/2\big]}\sqrt{-det(\tG_{ab}+e^{-\gamma\phi}\tB_{ab})}\,,\labell{dbiE}
\eeqa
To transform the D-brane action \reef{dbi3} to the Einstein frame, one needs the following relations between the two frames:
\beqa
\tR&\Longrightarrow&e^{-\gamma\phi}\bigg[\tR-\gamma p\tilde{\nabla}^2\phi-\frac{\gamma^2}{4} p(p-1)\prt_a\phi\prt^a\phi\bigg]\,,\nonumber\\
\Omega^{\mu}{}_{ab}&\Longrightarrow&\Omega^{\mu}{}_{ab}+\frac{\gamma}{2}\Big(\prt_a\phi\prt_bX^{\mu}+\prt_b\phi\prt_a X^{\mu}-\tG_{ab}\prt^{\mu}\phi\Big)\,,\nonumber\\
\Omega^{\mu}{}_{a}{}^a&\Longrightarrow&e^{-\gamma\phi}\bigg[\Omega^{\mu}{}_{a}{}^a+\frac{\gamma}{2}\Big(2\prt_a\phi\prt^aX^{\mu} -(p+1)\prt^{\mu}\phi\Big)\bigg]\,,
\eeqa
where we have also used  the relation   $\tG_{ab}\tG^{ba}=p+1$. 
Using the above transformations, one can find the   brane action corresponding to the bulk action \reef{bulkE4}. However, to find the D-brane action which is corresponding to the bulk action \reef{bulkE3}, we have to add to the D-brane action  those terms which are coming from the field redefinition \reef{red2}. Using the relation $\prt_a X^{\mu}\bot_{\mu\nu}=0$, one finds
\beqa
S_{D_p}^{(1)}&\!\!\!\!\!=\!\!\!\!\!& 
-\frac{ T_p}{2}\int d^{p+1}xe^{-\phi\big[1-\gamma(p-1)/2\big]}
\sqrt{-\tG }\bigg[\tR+\bot_{\mu\nu}\Big(\Omega^\mu{}_{ a}{}^a
\Omega^\nu{}_{ b}{}^b-{\Omega^{\mu}}_{ab}
\Omega^{\nu\,ab}\Big)\labell{dbiE3}\\
&&\qquad+e^{-2\gamma \phi}\Big(\frac{1}{24}H_{abc}H^{abc}+ \frac{1}{8} \bot^{\alpha\beta}
H_{\alpha\mu\nu}H_{\beta}{}^{\mu\nu}  -\frac{3}{8} \bot^{\alpha\beta} \bot^{\mu\nu}
H_{\alpha\mu\lambda}H_{\beta\nu}{}^{\lambda}\nonumber\\
&&\qquad+ \frac{5}{24} \bot^{\alpha\beta} \bot^{\mu\nu}\bot^{\lambda\sigma} H_{\alpha\mu\lambda}H_{\beta\nu\sigma}\Big) +\Big(\gamma(p+1)(\frac{\gamma p}{4}-1)+1\Big) \bot_{\mu\nu}\prt^{\mu}\phi \prt^{\nu}\phi\nonumber\\
 &&\qquad-(\gamma p-2)\bot_{\mu\nu}\Omega^{\mu}{}_a{}^a \prt^{\nu}\phi-\Big(3+\frac{\gamma^2}{4}p(p-1)\Big)\prt_a\phi\prt^a\phi -\gamma p\tilde{\nabla}^2\phi +\Delta  \bigg]\,,\nonumber
\end{eqnarray}
where  $\Delta$ is coming from the field redefinition \reef{red2} which is 
\beqa
\Delta&=&\frac{1}{2} \Big(1-\frac{\gamma}{2}(p+1)\Big)\Big(\gamma (\bot_{\mu\nu}\prt^{\mu}\phi\prt^{\nu}\phi+\prt_a\phi\prt^a\phi)+\frac{1}{12}H^2e^{-2\gamma\phi}\Big) \nonumber\\
&&+\frac{\gamma^2}{4}\Big( D(D-3)+4\Big)\prt_a\phi\prt^a\phi-\gamma\Big(\tilde{\nabla}^2\phi-\bot_{\mu\nu}
\Omega^{\mu}{}_a{}^a\prt^{\nu}\phi\Big)\,.  \nonumber
\eeqa
 Using the identity \reef{Hiden}, one can easily observe that the B-field couplings in the above action are exactly the couplings that have been found in \cite{Corley:2001hg} from the corresponding disk-level S-matrix elements.

  To check the dilaton couplings with the S-matrix elements, one may  use once more field redefinition. Since we have already used the field redefinition on the bulk fields to convert   the bulk action  to the standard form \reef{bulkE3}, we are not allowed anymore to use the field redefinition on the bulk fields. However, we can still use field redefinition on the brane fields. The variation of the DBI action \reef{dbiE} under $\bot_{\mu\nu}\delta X^{\nu}$ is
\beq
-T_p\int d^{p+1}x\, e^{-\phi\big[1-\gamma(p+1)/2\big]}\sqrt{-\tG}\bigg[\bot_{\mu\nu}\delta X^{\mu}\Big(-\big[1-\gamma(p+1)/2\big]\prt^{\nu}\phi-\Omega^{\nu}{}_a{}^a\Big)\bigg]\,.
\eeq
To convert the coupling with structure $\bot_{\mu\nu}
\Omega^{\mu}{}_a{}^a\prt^{\nu}\phi$ to the coupling  $\bot_{\mu\nu}\prt^{\mu}\phi\prt^{\nu}\phi$, one can use the following field redefinition:
\beqa
\delta X^{\mu}&=&-\frac{\alpha'}{2}\big(2+\gamma-\gamma p\big)e^{-\gamma\phi}\prt^{\mu}\phi\,.
\eeqa
Using also the integration by part to convert the coupling $\tilde{\nabla}^2\phi$ to the coupling with structure $\prt_a\phi\prt^a\phi$, one finds the following   result:
\beqa
S_{D_p}^{(1)} &\!\!\!\!\!=\!\!\!\!\!&
-\frac{ T_p}{2}\int d^{p+1}xe^{-\phi\big[1-\gamma(p-1)/2\big]}
\sqrt{-\tG }\bigg[\tR+\bot_{\mu\nu}\Big(\Omega^\mu{}_{ a}{}^a
\Omega^\nu{}_{ b}{}^b-{\Omega^{\mu}}_{ab}
\Omega^{\nu\,ab}\Big)\labell{dbiE4}\\
&&\qquad+e^{-2\gamma \phi}\Big(\frac{1}{24}H_{abc}H^{abc}+ \frac{1}{8} \bot^{\alpha\beta}
H_{\alpha\mu\nu}H_{\beta}{}^{\mu\nu}  -\frac{3}{8} \bot^{\alpha\beta} \bot^{\mu\nu}
H_{\alpha\mu\lambda}H_{\beta\nu}{}^{\lambda}\nonumber\\
&&\qquad+ \frac{5}{24} \bot^{\alpha\beta} \bot^{\mu\nu}\bot^{\lambda\sigma} H_{\alpha\mu\lambda}H_{\beta\nu\sigma}+\frac{1}{24}\big[1-\frac{\gamma}{2}(p+1)\big]H^2\Big)  \nonumber\\
 &&\qquad-\frac{\gamma^2}{16}\big[(D-2p-2)^2+2D-8\big] \Big(\prt_a\phi\prt^a\phi -\bot_{\mu\nu}\prt^{\mu}\phi\prt^{\nu}\phi\Big) \bigg]\,.\nonumber
\end{eqnarray}
This is the D-brane action which is corresponding to the bulk action \reef{bulkE3}. In fact, the action \reef{bulkE3} has been used in \cite{Corley:2001hg} to calculate the massless $t$-channel pole of the scattering amplitude of two  B-fields from D-brane. By subtracting this pole from the corresponding disk-level S-matrix elements at order $\alpha'$, the B-field   couplings have been found  in \cite{Corley:2001hg} which are exactly the same as the B-field couplings in above action.   The couplings of one graviton and one dilaton that have been found in \cite{Corley:2001hg} are also consistent with the above action. However, the couplings of two dilatons in above action are not exactly the same as the couplings that have been found in \cite{Corley:2001hg}. Using the   S-matrix element   calculated in \cite{Corley:2001hg},  we have reexamined the calculation of the couplings of two dilatons at $\alpha'$-order and we have found an exact agreement with the above action. The details of this calculation  appear in the Appendix A. This completes our illustration of perfect agreement between the manifestly T-duality invariant D-brane action \reef{dbi2} that we have found in this paper and the S-matrix calculations.

{\bf Acknowledgment}: This work is supported by Ferdowsi University of Mashhad under the grant P/259-1387/03/04.

\appendix
\section{\bf S-matrix element of two dilatons }

In this appendix we are going to show that the world volume  couplings of two dilatons in the Einstein frame that we have found in \reef{dbiE4} are reproduced by corresponding string theory S-matrix element.  
The disk-level scattering amplitude of two closed strings off a D-brane in bosonic string theory is given by the following expression \cite{Corley:2001hg}:
\beqa
 A&\sim&{d_1} B(-t/2,1+2s)+ {d_2} B(-t/2,2s)- {d_3} B(1-t/2,2s)+ {d_4} B(1-t/2,1+2s)\nonumber\\
&&+ {d_5} B(-1-t/2,1+2s)+ {d_6} B(1-t/2,-1+2s)+ {d_7} B(-1-t/2,-1+2s)\nonumber\\
&&- {d_8} B(-t/2,-1+2s)- {d_9} B(2-t/2,-1+2s)+ {d_{10}} B(3-t/2,-1+2s)\,,
\eeqa
where $t=-\alpha'k_{1\mu}k_2^{\mu}$  and $s=-\alpha' k_{1a}k_2^a/2$.  In above amplitude, $d_1,\,\cdots, d_{10}$ are some kinematic factors that depend on momenta and the polarization of the external states. We refer the interested readers to \cite{Corley:2001hg} for the explicit form of these factors. 

The dilaton amplitude     is given by replacing the   polarization tensors in $d$'s with the following expression:
\beqa
\veps_{\mu\nu}=\frac{1}{\sqrt{D-2}}(\eta_{\mu\nu}-k_{\mu}\ell_{\nu}-k_{\nu}\ell_{\mu})\,,
\eeqa
where the auxiliary vector  $\ell_{\mu}$  satisfying $\ell\inn k=1$, must be canceled in the whole amplitude. We have done this replacement and found the following result:
\beqa
A &\sim&(D-t-4) B\Big(-\frac{t}{2}-1,1+2 s\Big)+\frac{\Gamma (2 s-1) \Gamma (1-\frac{t}{2})}{4 t \Gamma (2 s-\frac{t}{2}+2)}\bigg[
\nonumber\\
&&+16 s^2  \Big[t \left(D-4 t \Tr V+t (t+7)+(\Tr V)^2-4\right)-4 \Big]\nonumber\\
&&-8 s t  \Big[D+t^2 (2-\Tr V)+t (1-\Tr V)^2+\Tr V
 (4-\Tr V)-8 \Big]\nonumber\\
&&+256 s^4 (t+1)-128 s^3 t (t-\Tr V+2)+(t-2) t^2 (2-\Tr V)^2\bigg]\,.
\eeqa
The auxiliary vector has been canceled, as expected. In above amplitude the matrix   $V_{\mu\nu}=-{\rm diag}(-1,1,\cdots,1,-1,-1,\cdots,-1)$.

To study the above amplitude at low energy, we have to expend it at low energy, \ie $\alpha'\rightarrow 0$. The result is 
\beqa
A&\sim&\frac{(2-\Tr V)^2t^2-4 s t (2-\Tr V)^2+16 s^2 (D-2)}{4 s t}\nonumber\\
&&+\frac{1}{2}(-4 s +t)[(\Tr V) ^2+2 D-8]+O(\alpha'^2)\,.\nonumber
\eeqa
The terms in the first line are at order $\alpha'^0$ which are reproduced by the couplings in the DBI action \reef{dbiE} and bulk action \reef{bulkE}, \cite{Corley:2001hg}. The terms in the second line are only contact terms which must be reproduced by the effective action at order $\alpha'$. It is easy to verify  that these terms are exactly reproduced by the dilaton couplings in the last line of \reef{dbiE4}.


\begin{thebibliography}{99}

\bibitem{Kikkawa:1984cp}
  K.~Kikkawa and M.~Yamasaki,
  Phys.\ Lett.\  B {\bf 149}, 357 (1984).
\bibitem{TB}  
T. Buscher, Phys. Lett. B  {\bf 194} (1987) 59; B {\bf 201} (1988) 466.

\bibitem{Giveon:1994fu}
  A.~Giveon, M.~Porrati and E.~Rabinovici,
  Phys.\ Rept.\  {\bf 244}, 77 (1994)
  [arXiv:hep-th/9401139].

\bibitem{Alvarez:1994dn}
  E.~Alvarez, L.~Alvarez-Gaume and Y.~Lozano,
  Nucl.\ Phys.\ Proc.\ Suppl.\  {\bf 41}, 1 (1995)
  [arXiv:hep-th/9410237].

\bibitem{Kugo:1992md} 
  T.~Kugo and B.~Zwiebach,
  Prog.\ Theor.\ Phys.\  {\bf 87}, 801 (1992)
  [hep-th/9201040].

\bibitem{Zwiebach:1992ie} 
  B.~Zwiebach,
  Nucl.\ Phys.\ B {\bf 390}, 33 (1993)
  [hep-th/9206084].


\bibitem{Witten:1985cc} 
  E.~Witten,
  Nucl.\ Phys.\ B {\bf 268}, 253 (1986).
 \bibitem{Hull:2009mi} 
  C.~Hull and B.~Zwiebach,
  JHEP {\bf 0909}, 099 (2009)  [arXiv:0904.4664 [hep-th]].  

\bibitem{Hohm:2010jy} 
  O.~Hohm, C.~Hull and B.~Zwiebach,
  JHEP {\bf 1007}, 016 (2010)  [arXiv:1003.5027 [hep-th]].  

\bibitem{Hohm:2010pp} 
  O.~Hohm, C.~Hull and B.~Zwiebach,
  JHEP {\bf 1008}, 008 (2010)  [arXiv:1006.4823 [hep-th]]. 


\bibitem{Metsaev:1987zx} 
  R.~R.~Metsaev and A.~A.~Tseytlin,
  Nucl.\ Phys.\ B {\bf 293}, 385 (1987).

\bibitem{Gross:1986mw} 
  D.~J.~Gross and J.~H.~Sloan,
  Nucl.\ Phys.\ B {\bf 291}, 41 (1987).  

\bibitem{Bachas:1999um} 
  C.~P.~Bachas, P.~Bain and M.~B.~Green,
  JHEP {\bf 9905}, 011 (1999)
  [hep-th/9903210].

\bibitem{Corley:2001hg} 
  S.~Corley, D.~A.~Lowe and S.~Ramgoolam,
  JHEP {\bf 0107}, 030 (2001)
  [hep-th/0106067].

\bibitem{Kaloper:1997ux} 
  N.~Kaloper and K.~A.~Meissner,
  Phys.\ Rev.\ D {\bf 56}, 7940 (1997)
  [hep-th/9705193].

\bibitem{Garousi:2009dj} 
  M.~R.~Garousi,
  JHEP {\bf 1002}, 002 (2010)
  [arXiv:0911.0255 [hep-th]].


\bibitem{Becker:2007zj}
  K.~Becker, M.~Becker and J.~H.~Schwarz,
  ``String theory and M-theory: A modern introduction,''
{\it  Cambridge, UK: Cambridge Univ. Pr. (2007) 739 p}

\bibitem{Meissner:1996sa} 
  K.~A.~Meissner,
  Phys.\ Lett.\ B {\bf 392}, 298 (1997)
  [hep-th/9610131].


\bibitem{Metsaev:1986yb} 
  R.~R.~Metsaev and A.~A.~Tseytlin,
  Phys.\ Lett.\ B {\bf 185}, 52 (1987).


\bibitem{Tseytlin:1986zz} 
  A.~A.~Tseytlin,
  Phys.\ Lett.\ B {\bf 176}, 92 (1986).


\bibitem{Leigh:1989jq}
  R.~G.~Leigh,
  Mod.\ Phys.\ Lett.\  A {\bf 4}, 2767 (1989).
  
 
\bibitem{Bachas:1995kx}
  C.~Bachas,
  Phys.\ Lett.\  B {\bf 374}, 37 (1996)
  [arXiv:hep-th/9511043].
  
\bibitem{Myers:1999ps} 
  R.~C.~Myers,
  JHEP {\bf 9912}, 022 (1999)
  [hep-th/9910053].


\bibitem{Ardalan:2002qt} 
  F.~Ardalan, H.~Arfaei, M.~R.~Garousi and A.~Ghodsi,
  Int.\ J.\ Mod.\ Phys.\ A {\bf 18}, 1051 (2003)
  [hep-th/0204117].

\bibitem{Garousi:2011fc} 
  M.~R.~Garousi,
  Phys.\ Lett.\ B {\bf 701}, 465 (2011)
  [arXiv:1103.3121 [hep-th]].

\bibitem{Meessen:1998qm}
  P.~Meessen and T.~Ortin,
  Nucl.\ Phys.\  B {\bf 541}, 195 (1999)
  [arXiv:hep-th/9806120].
  
\bibitem{Bergshoeff:1995as}
  E.~Bergshoeff, C.~M.~Hull and T.~Ortin,
  Nucl.\ Phys.\  B {\bf 451}, 547 (1995)
  [arXiv:hep-th/9504081].
\bibitem{Bergshoeff:1996ui}
  E.~Bergshoeff, M.~de Roo, M.~B.~Green, G.~Papadopoulos and P.~K.~Townsend,
  Nucl.\ Phys.\  B {\bf 470}, 113 (1996)
  [arXiv:hep-th/9601150].

\bibitem{Hassan:1999bv}
  S.~F.~Hassan,
  Nucl.\ Phys.\  B {\bf 568}, 145 (2000)
  [arXiv:hep-th/9907152].


\bibitem{Carter:1997pb} 
  B.~Carter,
  ``Brane dynamics for treatment of cosmic strings and vortons,''
  hep-th/9705172.
 

\bibitem{Peeters:2006kp} 
  K.~Peeters,
  Comput.\ Phys.\ Commun.\  {\bf 176}, 550 (2007)
  [cs/0608005 [cs.SC]].

\bibitem{Peeters:2007wn} 
  K.~Peeters,
  ``Introducing Cadabra: A Symbolic computer algebra system for field theory problems,''
  hep-th/0701238 [HEP-TH].



 
\end{thebibliography}
\end{document}